\documentclass[10pt,journal]{IEEEtran}
\ifCLASSINFOpdf
\else
   \usepackage[dvips]{graphicx}
\fi
\usepackage{amsmath}
\newtheorem{theorem}{Theorem}
\newtheorem{remark}{Remark}

\newtheorem{definition}{Definition}
\newtheorem{assumption}{Assumption}

\allowdisplaybreaks[3]

\begin{document}
%
\title{Distribution Modeling and Stabilization Control for Discrete-Time Linear Random Dynamical Systems Using
Ensemble Kalman Filter}
%
%
%

\author{Yohei~Hosoe 
        and~Dimitri~Peaucelle
\thanks{This work was partially
supported by JSPS KAKENHI Grant Number 17K14700.}
\thanks{Y.~Hosoe is with the Department of Electrical Engineering, 
        Kyoto University, Nishikyo-ku, Kyoto 615-8510, Japan
and with LAAS-CNRS, Univ.\ Toulouse, CNRS, Toulouse 31400, France
        (e-mail: hosoe@kuee.kyoto-u.ac.jp).}
\thanks{D.~Peaucelle is with LAAS-CNRS, Univ.\ Toulouse, CNRS, Toulouse 31400, France
(e-mail: peaucelle@laas.fr).}
}

\maketitle

\begin{abstract}
This paper studies an output feedback stabilization control framework for
discrete-time linear systems with stochastic dynamics determined
by an independent and identically distributed (i.i.d.) process.
The controller is constructed with an ensemble Kalman filter (EnKF) and
a feedback gain designed with our earlier result about state
feedback control.
The EnKF is also used for modeling the distribution behind the system,
which is required in the feedback gain synthesis.
The effectiveness of our control framework is demonstrated with
numerical experiments.
This study will become the first step toward the realization of learning
type control using our stochastic systems control theory.
\end{abstract}

\begin{IEEEkeywords}
Stochastic dynamics, 
distribution modeling, 
feedback control,
ensemble Kalman filter,
LMI optimization.
\end{IEEEkeywords}

%
\IEEEpeerreviewmaketitle

\section{Introduction}

The influence of stochastic noise in modeling and control is more or less inevitable in
practice.
In modeling of systems, for example, 
noise may cause a fluctuation of estimated
parameters.
Such a fluctuation is known to arise,
e.g., in flight vehicle systems \cite{Jategaonkar-book}.
If the influence is larger than we can disregard, it would be better to
take into account the presence of noise in modeling and control.
As a system class handling this issue,
this paper deals with discrete-time linear systems whose dynamics are determined
by a stochastic process (regarded as noise).
Such a system is called in the field of analytical dynamics a
discrete-time linear random dynamical system \cite{Arnold-book}.

Markov jump systems \cite{Costa-book} are
one of the most known special cases of random dynamical systems,
whose dynamics are determined by a finite-mode Markov
chain.
Although the class of stochastic dynamics that can be described with a
finite-mode Markov chain is limited from the viewpoint of possible noise, 
the study \cite{Costa-TAC14} extended the associated theory so that a more general
Markov process can be dealt with.
On the other hand, the authors are also developing another theoretical
framework of controlling random dynamical systems whose dynamics are
determined by a stochastic process that is independent and identically
distributed (i.i.d.) with respect to the discrete time
\cite{Hosoe-TAC18,Hosoe-TAC-sub}.
Since i.i.d.\ processes are a special case of Markov processes, one
might consider that our results could be covered by those for the Markov
jump systems.
However, this is not true and our studies have an advantage that we can
deal with unbounded supports for coefficient random matrices (depending
on an i.i.d.\ process) in the
system model, which are difficult to
deal with in the case of Markov jump systems.
The class of such coefficient random matrices includes those having, e.g.,
normally distributed entries.
If the process determining the system dynamics can be seen as an
i.i.d.\ process, and if we do not have sufficient prior information
on its support (i.e., the range of fluctuation),
then using our framework may be a good option.

In \cite{Hosoe-TAC18,Hosoe-TAC-sub}, we have discussed state
feedback controller synthesis for stabilization of the systems.
However, the situation where all the system states can be directly measured 
is considered to be unusual in practice.
In addition, the distribution of the stochastic process determining
system dynamics may be also unknown, which is required in designing
the state feedback gain with our theory.
The purpose of this paper is to resolve these issues by exploiting a
sequential Monte Carlo method called
the ensemble Kalman filter (EnKF) \cite{Evensen-EnKF,Evensen-book} in distribution
modeling as well as online state observation.
The EnKF can be used for solving nonlinear filtering problems.
The computational cost of the EnKF is relatively not expensive, and it
has been exploited in large-scale problems such as atmospheric data assimilation
\cite{houtekamer2016review}.
We have a hope that statistical methods such as the EnKF would be
compatible with our stochastic systems control theory.

This paper is organized as follows.
Section~\ref{sc:state-feedback} describes discrete-time linear
random dynamical systems whose dynamics are determined by an
i.i.d.~process, and then, briefly reviews the earlier result on
synthesis of a stabilizing state
feedback gain for the system.
Section~\ref{sc:EnKF} introduces a standard
usage of the EnKF in parameter estimation, and discusses an EnKF-based method of
modeling the
distribution behind the system, which is used for the above gain design.
Section~\ref{sc:control} proposes an output feedback control framework
using the EnKF and the feedback gain designed with the reviewed
result.
The effectiveness of such a framework is
demonstrated in Section~\ref{sc:num-exam} through numerical experiments.

We use the following notation in this paper.
The set of real numbers,
that of positive real numbers
and that of non-negative integers
are denoted by ${\bf R}$, ${\bf R}_+$ and ${\bf N}_0$, respectively.
The set of $n$-dimensional real
column vectors and that of $m\times n$ real matrices are denoted by
${\bf R}^n$ and ${\bf R}^{m\times n}$, respectively.
The set of 
$n\times n$ positive definite matrices is denoted by
${\bf S}^{n\times n}_{+}$.
The identity matrix of size $n$ is denoted by $I_n$. 
The Euclidean norm
is denoted by $||\cdot||$.
The vectorization of a matrix in the row
direction is denoted by ${\rm row}(\cdot)$, i.e., ${\rm row}(\cdot):=[{\rm row}_1(\cdot),\ldots,{\rm
row}_m(\cdot)]$, where $m$ is the number of rows of the matrix and ${\rm row}_i(\cdot)$ denotes the $i$th row.
The Kronecker product is denoted by $\otimes$.
The (block) diagonal matrix is denoted by ${\rm diag}(\cdot)$.
The Dirac delta function is denoted by $\delta(\cdot)$.
The expectation (i.e., the expected value) of a random variable is denoted by $E[\cdot]$; this notation
is also used for the expectation of a random matrix.
If $s$ is a random variable obeying the distribution ${\cal D}$,
then we represent it as $s \sim {\cal D}$.

\section{Discrete-Time Linear Random
Dynamical Systems and State Feedback Stabilization}
\label{sc:state-feedback}

\subsection{Discrete-Time Linear Random Dynamical Systems with
an i.i.d.~Process}
\label{ssc:sys}

Let us consider the $Z$-dimensional discrete-time stochastic process
$\xi=\left(\xi_k\right)_{k\in {\bf N}_0}$ satisfying the following assumption.

\begin{assumption}
\label{as:iid}
$\xi_k$ is independent and identically distributed (i.i.d.) with
 respect to the discrete time $k\in {\bf N}_0$.
\end{assumption}

This assumption naturally makes $\xi$ stationary and ergodic \cite{Klenke-book}.
The support of $\xi_k$ is denoted by ${\boldsymbol {\mit\Xi}}$.
By definition, ${\boldsymbol {\mit\Xi}} \subset {\bf R}^Z$, 
and ${\boldsymbol {\mit\Xi}}$ corresponds to the set of values
that $\xi_k$ can take (at each $k$).

With such a process $\xi$, consider the discrete-time linear random
dynamical system
\begin{equation}
x_{k+1} = A(\xi_k) x_k + B(\xi_k) u_k 
\label{eq:fr-sys}
\end{equation}
where $x_k \in {\bf R}^n$, $u_k \in {\bf R}^m$, 
$A:{\boldsymbol {\mit\Xi}} \rightarrow
{\bf R}^{n\times n}$, 
$B:{\boldsymbol {\mit\Xi}} \rightarrow
{\bf R}^{n\times m}$,
and the initial state $x_0$ is assumed to be
deterministic.
The representation of the coefficient random matrices in (\ref{eq:fr-sys})
is general in the sense that any random matrices (denoted by $A_k$ and
$B_k$) can be represented in the
form with appropriate $\xi$; we can always take $A(\cdot)$, $B(\cdot)$ and
$\xi$ such that $A(\xi_k)=A_k$ and $B(\xi_k)=B_k$ (under Assumption~\ref{as:iid}).
Our control approach is developed for such a system.

\subsection{Stability and State Feedback Stabilization}
\label{ssc:syn}

If full prior information about the distribution of $\xi_0$ as well as the structure of
$A(\cdot)$ and $B(\cdot)$ (involving constant coefficients) is available, and if the
exact value of the system state can be obtained at each time step, then we can readily use
our earlier results in \cite{Hosoe-TAC-sub} for stabilizing system (\ref{eq:fr-sys}).
This subsection briefly reviews the earlier results under such an
ideal situation,
before proceeding to the case with a lack of information.

Let us consider the state feedback
\begin{align}
&
u_k=F x_k \label{eq:state-feedback}
\end{align}
with the static time-invariant gain $F\in {\bf R}^{m\times n}$,
and the associated closed-loop system
\begin{align}
x_{k+1} = A_{\rm cl}(\xi_k) x_k,\ \ A_{\rm cl}(\xi_k)=A(\xi_k)+B(\xi_k)F.
\label{eq:closed-loop}
\end{align}
To define second-moment stability for this system, we introduce the
following assumption (which is actually a minimal requirement for the
definition, although the details are omitted).
\begin{assumption}
\label{as:bound}
The squares of elements of 
$A(\xi_k)$ and $B(\xi_k)$ 
are all Lebesgue integrable, i.e.,
\begin{align}
&
E[A_{ij}(\xi_k)^2]<\infty,\ \ 
E[B_{ij}(\xi_k)^2]<\infty,
\label{eq:as-bound}
\end{align}
where $A_{ij}(\xi_k)$ and $B_{ij}(\xi_k)$ represent the $(i,j)$-entries
of $A(\xi_k)$ and $B(\xi_k)$, respectively.
\end{assumption}

For each fixed $F$, this assumption ensures the squares of elements of
$A_{\rm cl}(\xi_k)$ are also Lebesgue integrable.
Under Assumptions~\ref{as:iid} and \ref{as:bound}, we define exponential second-moment stability
(i.e., exponential mean square stability) \cite{Kozin-Auto69} as follows.

\begin{definition}
\label{df:expo}
The system (\ref{eq:closed-loop}) with a fixed $F$ satisfying
Assumptions~\ref{as:iid} and \ref{as:bound}
 is said to be exponentially stable
 in the second moment if there exist $a\in {\bf R}_+$ and $\lambda \in
 (0,1)$ such that
\begin{align}
&
\sqrt{E[||x_k||^2]} \leq a ||x_0|| \lambda^k\ \ \ (\forall k \in {\bf
N}_0, \forall x_0 \in {\bf R}^n).
\label{eq:exp-def}
\end{align}
\end{definition}

This stability notion can be characterized by a Lyapunov inequality as
in the following theorem \cite{Hosoe-TAC-sub}.

\begin{theorem}
\label{th:lyap}
Suppose the open-loop system (\ref{eq:fr-sys}) satisfies Assumptions~\ref{as:iid}
and \ref{as:bound}.
For given $F\in {\bf R}^{m\times n}$,
the following two conditions are equivalent.
\begin{enumerate}
 \item 
The closed-loop system (\ref{eq:closed-loop}) is exponentially stable in the second moment.
\item
There exist $P\in {\bf S}^{n\times n}_+$ and $\lambda\in (0,1)$ such
 that
\begin{align}
&
E[\lambda^2 P - A_{\rm cl}(\xi_0)^T P A_{\rm cl}(\xi_0)]\geq 0.\label{eq:lyap-lambda}
\end{align}
\end{enumerate}
\end{theorem}

Based on the Lyapunov inequality (\ref{eq:lyap-lambda}), we can further obtain the
following theorem \cite{Hosoe-TAC-sub}, which gives an inequality condition for
designing a stabilizing state feedback gain.

\begin{theorem}
\label{th:syn}
Suppose the open-loop system (\ref{eq:fr-sys}) satisfies Assumptions~\ref{as:iid}
and \ref{as:bound}.
There exists a gain $F$ such that the closed-loop system
(\ref{eq:closed-loop}) is exponentially stable in the second moment
if and only if there exist $X\in {\bf S}^{n\times n}_+$,
$Y\in {\bf R}^{m\times n}$  and $\lambda\in (0,1)$ satisfying
\begin{align}
&
\begin{bmatrix}
\lambda^2 X& \ast\\
\bar{G}^\prime_{A}X+\bar{G}^\prime_{B} Y &
X\otimes I_{\bar{n}}
\end{bmatrix}\geq 0 \label{eq:lmi-syn}
\end{align}
for $\bar{G}^\prime_{A}$ and $\bar{G}^\prime_{B}$ given by

\begin{align}
&
\bar{G}^\prime_A :=[\bar{G}_{A1}^T, \ldots,
 \bar{G}_{An}^T]^T \in {\bf R}^{n\bar{n} \times n},\label{eq:def-GpA}\\
&
\bar{G}^\prime_B :=[\bar{G}_{B1}^T, \ldots,
\bar{G}_{Bn}^T]^T \in {\bf R}^{n\bar{n} \times m},
\label{eq:def-GpB}\\
&
\bar{G}=:
\left[\bar{G}_{A1}, \ldots, \bar{G}_{An}, \bar{G}_{B1}, \ldots,
\bar{G}_{Bn}\right] \notag\\
& 
(\bar{G}_{Ai} \in {\bf R}^{\bar{n}\times n}, \bar{G}_{Bi} \in {\bf
R}^{\bar{n}\times m}\ (i=1,\ldots,n))
\end{align}
with a matrix $\bar{G} \in {\bf R}^{\bar{n}\times(n+m)n}\ (\bar{n}\leq (n+m)n)$ satisfying
\begin{align}
&
\bar{G}^T \bar{G}=E\big[[{\rm row}(A(\xi_0)), {\rm row}(B(\xi_0))]^T
\notag \\
&\cdot [{\rm row}(A(\xi_0)), {\rm row}(B(\xi_0))]\big].
\label{eq:equiv-rep-decom-syn}
\end{align}
In particular, $F=YX^{-1}$ is one such stabilizing gain.
\end{theorem}

While the Lyapunov inequality (\ref{eq:lyap-lambda}) involved decision variables
contained in the expectation operation, 
the decision variables in (\ref{eq:lmi-syn}) are all uncontained in the
expectation operation.
Hence, once $\bar{G}^\prime_{A}$ and $\bar{G}^\prime_{B}$ are
calculated, (\ref{eq:lmi-syn}) can be solved as a standard
linear matrix inequality (LMI) (for each fixed $\lambda$).
By minimizing $\lambda$ through a bisection with respect to
$\lambda^2$ under such an LMI, we can obtain a stabilizing gain that is optimal in the
sense of exponential second-moment stability; the minimal $\lambda$
corresponds to that in (\ref{eq:exp-def}), which corresponds to the
convergence rate of $\sqrt{E[||x_k||^2]}$ with respect to $k$.

As reviewed above, our earlier results can be readily used when 
full information of system (\ref{eq:fr-sys}) (except the value of $\xi_k$) is available.
However, it is not always possible to access full information in practical
control problems, and we may have to achieve some required performance
only with limited available information.
To exploit our control theory even in such a situation,
the following sections consider using the ensemble Kalman filter
\cite{Evensen-EnKF,Evensen-book} in modeling of unknown system parts as well as
online estimation of the system state.

\section{Distribution Modeling Using Ensemble Kalman Filter}
\label{sc:EnKF}

\subsection{Ensemble Kalman Filter}
\label{ssc:EnKF}

The ensemble Kalman filter (EnKF) is a sequential Monte Carlo method
that solves nonlinear filtering problems.
The problem of estimating system states and coefficient matrix parameters for
deterministic linear systems can be seen a special case of
nonlinear filtering problems, and the EnKF is known to be useful for such estimation.
Before proceeding to the arguments for our random dynamical systems,
let us briefly review a standard usage of the EnKF in this subsection (see,
e.g., \cite{Evensen-book} for further details).

As a simulation model in the EnKF, we consider the state equation
\begin{align}
&
\psi_k = f_k(\psi_{k-1}) + w_k
\label{eq:sim-state}
\end{align}
and the observation equation
\begin{align}
&
z_k = h_k(\psi_k) + v_k
\label{eq:sim-observ}
\end{align}
with given nonlinear functions $f_k(\cdot)$ and $h_k(\cdot)$, where $\psi_k$, 
$z_k$, $v_k$ and $w_k$ are the state, the observation,
the system noise and the observation noise, respectively.
The noise inputs are assumed to be independent of each other and obey the following multi-dimensional
zero-mean normal distributions with
given covariance matrices $Q$ and $R$.
\begin{align}
&
w_k \sim {\cal N}(0,Q),\ \ v_k \sim {\cal N}(0,R)
\label{eq:sim-noise-dist}
\end{align}
In the filtering problem, $\psi_k$ is regarded as a random vector, which
corresponds to an estimate of the system state and unknown
parameters\footnote{Unknown parameters are also dealt with as the state
of the simulation model in the parameter estimation case.} of a
plant.
In particular, the EnKF uses a set 
$\{\psi^{(i)}_{k| k}\}_{i=1,\ldots,M}$
of sample values called an
ensemble for approximating the distribution of $\psi_k$; each ensemble
member is simulated with the above simulation model.
In this paper, we consider
approximating the density function $p(\psi_k | {\bf y}_{1:k-1})$ of
the prior probability distribution as
\begin{align}
&
p(\psi_k | {\bf y}_{1:k-1}) \simeq
\frac{1}{M}\sum_{i=1}^{M}\delta(\psi_k - \psi^{(i)}_{k| k-1}),\label{eq:as-prior}
\end{align}
where ${\bf y}_{1:k}:=(y_i)_{i=1,\ldots,k}$ is the measurement (i.e.,
the output of the real plant) up to $k$, and
$\psi^{(i)}_{k| k-1}$ is the forecast of state $\psi_k$ 
calculated with the (filtered) $i$th ensemble member $\psi^{(i)}_{k-1|
k-1}$ at $k-1$.
Under the approximation (\ref{eq:as-prior}),
an EnKF is given as follows, where $w_k^{(i)}$ and $v_k^{(i)}\
(i=1,\ldots,M; k\in {\bf N})$ are
independent and their distributions are given in a
fashion similar to (\ref{eq:sim-noise-dist}) (with common $Q$ and $R$ for $i$).

\medskip
\noindent
{\it State prediction (state update)}:
\begin{align}
&
\psi_{k| k-1} = \frac{1}{M}\sum_{i=1}^{M} \psi^{(i)}_{k| k-1},\label{eq:state-pred1}\\
&
\psi^{(i)}_{k| k-1} = f_k(\psi^{(i)}_{k-1| k-1}) + w_k^{(i)}.\label{eq:state-pred2}
\end{align}

\medskip
\noindent
{\it Output prediction (output update)}:
\begin{align}
&
z_k = \frac{1}{M}\sum_{i=1}^{M} z^{(i)}_k,\label{eq:output-pred1}\\
&
z^{(i)}_k = h_k(\psi^{(i)}_{k| k-1}) + v_k^{(i)}.\label{eq:output-pred2}
\end{align}

\medskip
\noindent
{\it Filtering}:
\begin{align}
&
\psi^{(i)}_{k| k}=\psi^{(i)}_{k| k-1}+K_k(y_k - z^{(i)}_k),\label{eq:filtering}\\
& 
K_k = U_k(V_k)^{-1},\label{eq:kalman-gain}\\
&
U_k = \frac{1}{M-1}\sum_{i=1}^{M}(\psi^{(i)}_{k| k-1} - \psi_{k| k-1})
(z^{(i)}_k - z_k)^T,\label{eq:kalman-u}\\
&
V_k = \frac{1}{M-1}\sum_{i=1}^{M}(z^{(i)}_k - z_k)
(z^{(i)}_k - z_k)^T.\label{eq:kalman-v}
\end{align}

\medskip

In the above EnKF, $K_k$ corresponds to a Kalman gain.
With this filter, the ensemble $\{\psi^{(i)}_{k| k}\}_{i=1,\ldots,M}$
approximates the density function $p(\psi_k | {\bf y}_{1:k})$ of the
posterior probability distribution as
\begin{align}
&
p(\psi_k | {\bf y}_{1:k}) \simeq
\frac{1}{M}\sum_{i=1}^{M}\delta(\psi_k - \psi^{(i)}_{k| k}).
\label{eq:post-dist}
\end{align}
Through a sequential use of (\ref{eq:state-pred1})--(\ref{eq:kalman-v})
with respect to $k$, the above posterior
distribution gives an estimate of the system state and unknown parameters.

\begin{remark}
In this subsection, we introduced a standard usage of the EnKF.
However, this is not a unique usage, and there are varieties; 
using EnKFs is also not a unique option for nonlinear filtering.
For more details about the varieties and other approaches, see
\cite{Evensen-EnKF,Evensen-book,Stordal-2011} and other sophisticated articles.
Since discussing the differences among them is beyond the scope of this
paper, we only deal with the above type of EnKF.
\end{remark}

\subsection{Distribution Modeling}
\label{ssc:est}

In most practical control problems, the state of a plant cannot be directly
measured, and some observation method is needed when we use state
feedback controllers.
The same comment also applies to our random dynamical systems, as stated
in Section~\ref{sc:state-feedback}.
Hence, we consider the following output equation in addition to the
state equation (\ref{eq:fr-sys}) for the system model\footnote{We
consider two models of a plant in this paper.
One is the simulation model used in the EnKF and the other is the system model for
representing a plant in controller synthesis.}.
\begin{align}
&
y_k = C(\xi_k)x_k + D(\xi_k) u_k
\label{eq:fr-sys-output}
\end{align}
We assume that $x_k$ (and $\xi_k$) cannot be measured directly, and only $y_k$ is
available at each $k$.
The goal of this paper is to control the plant modeled by (\ref{eq:fr-sys}) and
(\ref{eq:fr-sys-output}) with our control approach reviewed in Section~\ref{sc:state-feedback}
and an EnKF.
In particular, we consider the situation where the prior
information on the distribution of $\xi_k$ is unavailable, 
and regard $\xi_k$ as a part of the state in the simulation model
(\ref{eq:sim-state}) and (\ref{eq:sim-observ}), i.e.,
\begin{align}
&
\psi_k=[x_k^T, \xi_k^T]^T.\label{eq:psi-x-xi}
\end{align}
This is reasonable since the prior information on the accurate distributions of 
fluctuating parameters is less usual to be provided
than that on constant parameters in practical problems.
To design a gain $F$ in (\ref{eq:state-feedback}) in such
a situation, this subsection first uses the EnKF for modeling the
distribution of $\xi_k$.

In this subsection, we temporarily see $u_k$ as an input for identification.
To reflect the influence of $u_k$ also in the simulation model of the EnKF,
we consider in (\ref{eq:state-pred2}) and (\ref{eq:output-pred2}) 
$f_k(\cdot)$ and $h_k(\cdot)$ such that
%
%
\begin{align}
&
f_k(\psi_{k-1})=f(\psi_{k-1}, u_{k-1})\notag \\
&\hspace{14mm}
=
\begin{bmatrix}
A(\xi_{k-1}) x_{k-1} + B(\xi_{k-1}) u_{k-1}\\
\xi_{k-1}
\end{bmatrix},\label{eq:f-A-B}\\
&
h_k(\psi_k)=h(\psi_{k}, u_k)=C(\xi_k)x_k + D(\xi_k) u_k\label{eq:h-C-D}
\end{align}
under (\ref{eq:psi-x-xi}).
Obviously, these functions are nonlinear in $\psi_k$ (even when $A$, $B$, $C$ and
$D$ are linear in $\xi_k$).
Here, if $\xi_k$ is supposed to be an unknown deterministic parameter
vector, the usual deterministic parameter estimation can be carried out
through the corresponding EnKF.
In practical parameter estimation problems, however, the estimated value of a
parameter more or less fluctuates even after sufficient time has elapsed. 
If the fluctuation is larger than we can disregard,
it might be better to take account of it in modeling and control.
This is why we deal with representation (\ref{eq:fr-sys}) and
(\ref{eq:fr-sys-output}) with random $\xi_k$ as the system model.

Under Assumption~\ref{as:iid} on $\xi$,
the time sequence of the estimated value of $\xi_k$ obtained by the EnKF could be seen
as giving an empirical distribution of $\xi_k$.
This is our basic idea of modeling random $\xi_k$.
To make the point clearer, we introduce the notation $\xi_{k| k}^{(i)}$
for representing the parameter part of the filtered $i$th ensemble member 
$\psi_{k| k}^{(i)}$ defined in a way similar to (\ref{eq:psi-x-xi}).
By inputting a signal for identification such as maximal length
sequences to a plant (modeled by 
system (\ref{eq:fr-sys}) and
(\ref{eq:fr-sys-output})) as $u_k$,
the EnKF connected to the plant provides us with the time sequence of $\xi_{k| k}^{(i)}$ (for
$i=1,\ldots, M$).
Then, one of the reasonable estimates of $\xi_k$ can be given by
\begin{align}
&
\xi_{k| k} = \frac{1}{M}\sum_{i=1}^M \xi_{k| k}^{(i)},
\end{align}
which corresponds to the mean of the parameter-related part of the posterior
distribution (\ref{eq:post-dist}).
If $\xi_{k| k}$ can be seen as being i.i.d.\ with respect to $k$ in
association with Assumption~\ref{as:iid},
the set $\{\xi_{k| k}\}_{k=K_0,\ldots,K_0+K_1}$ with
sufficiently large integers $K_0$ and $K_1$
corresponds to an empirical distribution of the estimate of $\xi_k$.
Although
obtaining the exact value of $\xi_k$ at each $k$ is generally
difficult even with EnKFs, we
have a hope that the obtained empirical distribution would not be
totally different from the
original distribution (if such an original actually exists), and 
decided to employ it as the modeling result of the distribution of
$\xi_k$.
If we have the prior information about the class of the distribution of
$\xi_k$ (such as being a normal distribution), the empirical distribution could
be further exploited for determining its hyperparameters.

\begin{remark}
Not only the information about the distribution with respect to $k$ but
also that with respect to $i$ for $\xi_{k| k}^{(i)}$ might be useful in modeling the
distribution of $\xi_k$.
However, using only the information about $i$ at some fixed $k$ is
unreasonable since $\xi_{k| k}^{(i)}\ (i=1,\ldots,M)$ are distributed
mainly because of the artificial noises $w_k^{(i)}$ and
$v_k^{(i)}$ in the EnKF.
\end{remark}

\begin{remark}
Parameter estimation is usually carried out for stable systems.
Hence, the plant in this section basically has to be stable
(in a stochastic sense).
One might consider this contradicts our tackling a stabilization
problem.
However, this is a sort of dilemma that is in common with conventional
deterministic systems modeling and control.
In most practical control problems, we usually have to first stabilize
plants without their models by using, e.g., PID controllers,
if they seem to be unstable.
After such stabilization, we identify the systems, and design controllers with
sophisticated model-based control theories to achieve required
performance.
A similar comment also applies to our problem.
Although this paper is the first attempt at combining 
a nonlinear filtering approach with our recent results,
and hence, only the convergence rate $\lambda$ can be dealt
with as the control performance index at this moment,
other performance indices such as $H_2$ norm
may be available in the future, as is the case with deterministic
systems control \cite{Boyd-book}.
\end{remark}

\begin{figure*}[t]
\centering
\begin{picture}(12.5,6.5)(0,0)
\put(1,1){\vector(0,1){1.5}}
\put(0,2.5){\framebox(2,1.5){\parbox[c]{1.5cm}{Plant}}}
\put(1,4){\line(0,1){1.5}}
\put(0.8,4.75){\makebox(0,0)[r]{$y_k$}}
\put(1,5.5){\vector(1,0){4}}
\put(5,4.5){\framebox(2,1.5){\parbox[c]{1.5cm}{Fitering}}}
\put(7,5.25){\vector(1,0){2}}
\put(8,5.45){\makebox(0,0)[b]{$\{\psi^{(i)}_{k|k}\}$}}
\put(9,4.5){\framebox(2,1.5){\parbox[c]{1.5cm}{State\\ prediction}}}
\put(10,4.5){\line(0,-1){0.75}}
\put(10.2,4.125){\makebox(0,0)[l]{$\{\psi^{(i)}_{k|k-1}\}$}}
\put(10,3.75){\circle*{0.08}}
\put(10,3.75){\vector(0,-1){0.75}}
\put(9,1.5){\framebox(2,1.5){\parbox[c]{1.5cm}{Control\\ input\\ decision}}}
\put(9,2.25){\line(-1,0){1}}
\put(8.5,2.05){\makebox(0,0)[t]{$u_k$}}
\put(8,2.25){\circle*{0.08}}
\put(8,2.25){\vector(-1,0){1}}
\put(5,1.5){\framebox(2,1.5){\parbox[c]{1.5cm}{Output\\ prediction}}}
\put(5,2.25){\line(-1,0){1}}
\put(4.5,2.05){\makebox(0,0)[t]{$\{z_k^{(i)}\}$}}
\put(4,2.25){\line(0,1){3}}
\put(4,5.25){\vector(1,0){1}}

\put(10,3.75){\line(-1,0){4}}
\put(6,3.75){\circle*{0.08}}
\put(6,3.75){\vector(0,1){0.75}}

\put(6,3.75){\vector(0,-1){0.75}}

\put(8,2.25){\line(0,-1){1.25}}
\put(8,1){\circle*{0.08}}
\put(8,1){\line(1,0){4}}
\put(12,1){\line(0,1){4.5}}
\put(12,5.5){\vector(-1,0){1}}

\put(8,1){\line(-1,0){7}}
\put(3.5,0){\line(1,0){0.05}} \multiput(3.55,0)(0.2,0){44}{\line(1,0){0.1}} \put(12.35,0){\line(1,0){0.15}}
\put(8,0.2){\makebox(0,0)[b]{Controller with EnKF}}
\put(12.5,0){\line(0,1){0.1}} \multiput(12.5,0.1)(0,0.2){31}{\line(0,1){0.1}} \put(12.5,6.3){\line(0,1){0.2}}
\put(12.5,6.5){\line(-1,0){0.05}} \multiput(12.45,6.5)(-0.2,0){44}{\line(-1,0){0.1}} \put(3.65,6.5){\line(-1,0){0.15}}
\put(3.5,6.5){\line(0,-1){0.1}} \multiput(3.5,6.4)(0,-0.2){31}{\line(0,-1){0.1}} \put(3.5,0.2){\line(0,-1){0.2}}

\end{picture}
\caption{Signal flows in our controlled system.}
\label{fig:wcontroller}
\end{figure*}
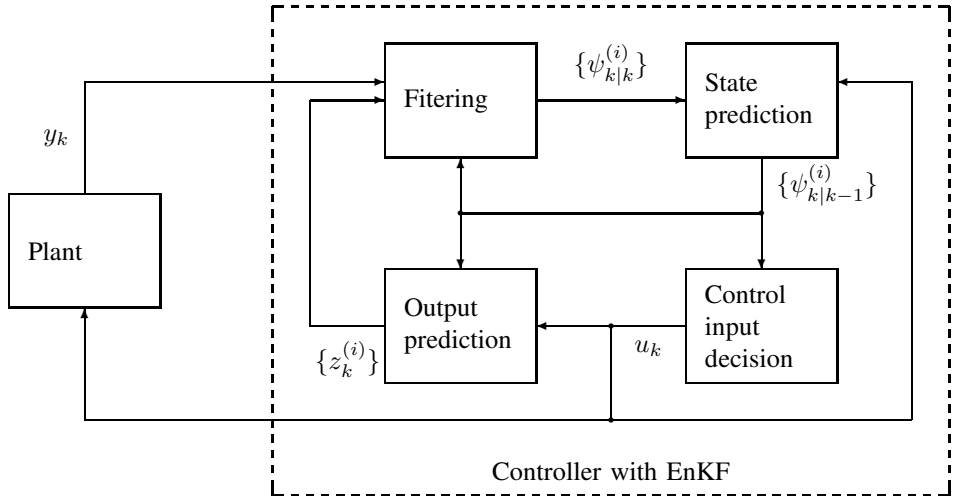

\section{Closed-Loop System with Ensemble Kalman Filter and
Stabilization Control}
\label{sc:control}

With the modeling result of the distribution of $\xi_k$ in the preceding
section, 
the expectation in (\ref{eq:equiv-rep-decom-syn}) can be calculated.
That is, we have
\begin{align}
\bar{G}^T \bar{G}=&\frac{1}{K_1+1}\sum_{k=K_0}^{K_0+K_1} ([{\rm row}(A(\xi_{k|k})), {\rm row}(B(\xi_{k|k}))]^T
\notag \\
&\cdot [{\rm row}(A(\xi_{k|k})), {\rm row}(B(\xi_{k|k}))]).\label{eq:sample-mean}
\end{align}
Hence, a state feedback gain $F$ that minimizes $\lambda$ in
(\ref{eq:exp-def}) for the {\it ideal} closed-loop system (\ref{eq:closed-loop}) with such modeled $\xi_k$ can be
obtained by minimizing $\lambda$ with respect to the inequality condition in
Theorem~\ref{eq:state-feedback}. 
Although the state feedback gain cannot be directly used for the
state-unmeasurable system (\ref{eq:fr-sys}) and
(\ref{eq:fr-sys-output}), 
it may work well if the state is observed (i.e., sequentially estimated) from the measurable output
with satisfactory accuracy\footnote{The principle of separation of estimation
and control has not been confirmed to hold for our problem at this moment.}.
The EnKF can be used also as a state observer, and this section discusses
the closed-loop control framework using it.

The algorithm of the EnKF for state observation is similar to that described
in the preceding section.
Only an essential difference is that we have to carry out online
estimation of the state in the closed-loop control, which was needless
in the preceding offline modeling. 
In the modeling, we used (the $\xi$-related part of) the filtered ensemble members $\psi_{k|
k}^{(i)}$ for obtaining the distribution model of $\xi_k$.
The filtered ensemble members are, however, calculated with $y_k$ as in
(\ref{eq:filtering}), which is the output of the plant at $k$.
Since this $y_k$ is the result of inputting $u_k$ on the plant at $k$,
and since $u_k$ has to be determined with (an estimate of) the state at
$k$, using (the $x$-related part of) $\psi_{k| k}^{(i)}$ for
online estimation is generally difficult.
Hence, we use the forecasts $\psi_{k| k-1}^{(i)}\ (i=1,\ldots,M)$ for online estimation
of $x_k$.
More precisely, we regard
\begin{align}
&
x_{k| k-1} = \frac{1}{M}\sum_{i=1}^M x_{k| k-1}^{(i)} \label{eq:x-predic}
\end{align}
as the estimate of $x_k$, where $x_{k| k-1}^{(i)}$ denotes the $x$-related part of $\psi_{k|
k-1}^{(i)}$, i.e.,
\begin{align}
&
\psi_{k| k-1}^{(i)}
=
\left[x_{k| k-1}^{(i)T}, \xi_{k| k-1}^{(i)T}\right]^T.\label{eq:psi-x-xi-i}
\end{align}
Here, since we already obtained a model of the distribution of $\xi_k$,
one might consider it needless to estimate $\xi_k$
simultaneously with $x_k$ as in (\ref{eq:psi-x-xi-i}).
However, we deal with this $\psi_{k| k-1}^{(i)}$ even in the present state
observation.
This is mainly because we cannot obtain the exact value of $\xi_k$
in the simulation model (\ref{eq:sim-state}) and (\ref{eq:sim-observ}) even if
we take a sample from the distribution model at each
$k$; what is worse, the sample
is generally independent of true $\xi_k$, which might cause an unnecessary
deterioration of the accuracy of the estimation of $x_k$.
Another important reason is the hedge against possible modeling errors.

With the above idea of state observation, our control framework with
an EnKF can be given as follows.

\medskip
\noindent
{\it State prediction}: (\ref{eq:state-pred1}) and (\ref{eq:state-pred2}) with
$f_k(\cdot)=f(\cdot,u_{k-1})$ such that (\ref{eq:psi-x-xi}) and (\ref{eq:f-A-B}).

\medskip
\noindent
{\it Control input decision}: (\ref{eq:x-predic}), (\ref{eq:psi-x-xi-i}) and
\begin{align}
&
u_k = F x_{k| k-1}.\label{eq:sf-enkf}
\end{align}

\medskip
\noindent
{\it Output prediction}: (\ref{eq:output-pred1}) and (\ref{eq:output-pred2}) with
$h_k(\cdot)=h(\cdot,u_k)$ such that (\ref{eq:psi-x-xi}) and (\ref{eq:h-C-D}).

\medskip
\noindent
{\it Filtering}: (\ref{eq:filtering})--(\ref{eq:kalman-v}).

\medskip

To facilitate further understanding, we also provide Fig.~\ref{fig:wcontroller} for showing
the signal flows between these four functions (constituting the whole controller)
as well as their connection with the plant.
Since the computational cost of the EnKF is not expensive compared to other
heuristic nonlinear filtering approaches, and since the internal control law
(\ref{eq:sf-enkf}) is very simple, this controller would work with a
practical sampling rate in various control problems.
In addition, if the EnKF provides a satisfactorily accurate estimate of the state
online, the closed-loop system is
expected to be stabilized by Theorem~\ref{th:syn}.

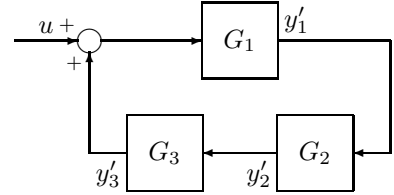
\begin{figure}[t]
\centering
\begin{picture}(5,2.5)(0,0)
\put(0,2){\vector(1,0){0.85}}
\put(0.425,2.1){\makebox(0,0)[b]{$u$}}
\put(0.7,2.2){\makebox(0,0){$ \scriptstyle + $}}
\put(1,2){\circle{0.3}}
\put(1.15,2){\vector(1,0){1.35}}
\put(2.5,1.5){\framebox(1,1){$G_1$}}
\put(3.5,2){\line(1,0){0.5}}
\put(3.75,2.1){\makebox(0,0)[b]{$y_1^\prime$}}
\put(4,2){\line(1,0){1}}
\put(5,2){\line(0,-1){1.5}}
\put(5,0.5){\vector(-1,0){0.5}}
\put(3.5,0){\framebox(1,1){$G_2$}}
\put(3.5,0.5){\line(-1,0){0.5}}
\put(3.25,0.4){\makebox(0,0)[t]{$y_2^\prime$}}
\put(3,0.5){\vector(-1,0){0.5}}
\put(1.5,0){\framebox(1,1){$G_3$}}
\put(1.5,0.5){\line(-1,0){0.5}}
\put(1.25,0.4){\makebox(0,0)[t]{$y_3^\prime$}}
\put(1,0.5){\vector(0,1){1.35}}
\put(0.8,1.7){\makebox(0,0){$ \scriptstyle + $}}
\end{picture}
\caption{Plant consisting of three subsystems.}
\label{fig:network}
\end{figure}

\section{Numerical Experiments}
\label{sc:num-exam}

\subsection{Application to Partially Unknown System}

\begin{figure*}[t]
\hspace{-18mm}
  \includegraphics[width=1.2\linewidth]{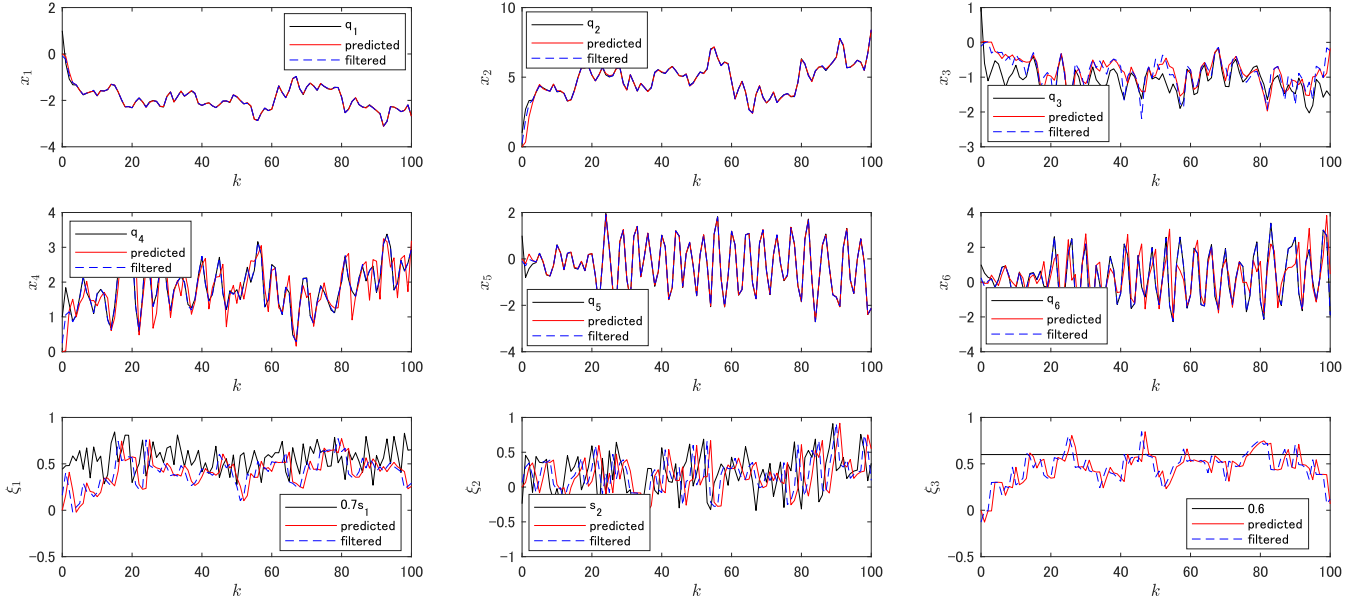}
  \caption{States and parameters of open-loop plant observed with EnKF.}
  \label{fig:estimation}
\end{figure*}

Consider the three subsystems described by
\begin{align}
&
G_i:
\begin{cases}
x^{\prime}_{i,k+1}=A_{ik} x^{\prime}_{ik}+B_{ik} u^{\prime}_{ik}\\
y^{\prime}_{ik}=C_{ik} x^{\prime}_{ik}
\end{cases}\ \ (i=1,2,3)\\
&
A_{1k}=
\begin{bmatrix}
0 & -0.4\\
1 & 1.3
\end{bmatrix},\ \ 
A_{2k}=
\begin{bmatrix}
0 & -0.6\\
1 & 0.7 s_{1k}
\end{bmatrix},\notag\\
& 
A_{3k}=
\begin{bmatrix}
0 & -0.8\\
1 & s_{2k}
\end{bmatrix},\ \ 
B_{1k}=
\begin{bmatrix}
0 \\ 0.4
\end{bmatrix},\ \ 
B_{2k}=
\begin{bmatrix}
0 \\ 0.4
\end{bmatrix},\notag\\
&
B_{3k}=
\begin{bmatrix}
0 \\ s_{2k}
\end{bmatrix},\ \ 
C_{ik}=
\begin{bmatrix}
0 & 1
\end{bmatrix}\ (i=1,2,3),
\end{align}
where $s_1$ and $s_2$ are i.i.d.~processes that
are independent of each other and distributed as
\begin{align}
&
s_{1k} \sim N(0.8, 0.2^2),\ \ s_{2k} \sim N(0.2, 0.3^2)
\end{align}
for each $k\in {\bf N}_0$.
Let us further consider the networked system described in Fig.~\ref{fig:network} consisting of
the above subsystems, where $u$ is the input and
$y_k=[y^{\prime}_{1k}, y^{\prime}_{2k}, y^{\prime}_{3k}]^T$ is the
measurable output.
This networked system is dealt with as the plant to be controlled in this section.
With $q_k=[x^{\prime T}_{1k}, x^{\prime T}_{2k}, x^{\prime
T}_{3k}]^T$, the plant can be described as follows.
\begin{align}
& q_{k+1}=
\begin{bmatrix}
 A_{1k} & 0 & B_{1k}C_{3k}\\
B_{2k}C_{1k} & A_{2k} & 0\\
0 & B_{3k}C_{2k} & A_{3k}
 \end{bmatrix}q_{k}
+
\begin{bmatrix}
 B_{1k}\\ 0\\ 0
 \end{bmatrix}u_{k},\label{eq:co-state}\\
&
y_k = {\rm diag}(C_{1k},C_{2k},C_{3k}) q_{k}\label{eq:co-output}
\end{align}

For the above plant,
we assume the situation where its prior information
is partially unknown and the system model is given by
(\ref{eq:fr-sys}) and (\ref{eq:fr-sys-output}) with coefficients
\begin{align}
&
A(\xi_k)=
\begin{bmatrix}
0 & -0.4 & 0 & 0 & 0 & 0\\
1 & 1.3 & 0 & 0 & 0 & 0.4\\
0 & 0 & 0 & -\xi_{3k} & 0 & 0 \\
0 & 0.4 & 1 & \xi_{1k} & 0 & 0\\
0 & 0 & 0 & 0 & 0 & -0.8\\
0 & 0 & 0 & \xi_{2k} & 1 & \xi_{2k}
\end{bmatrix},\notag\\
&
B(\xi_k)=
\begin{bmatrix}
0 & 0.4 & 0 & 0 & 0 & 0
\end{bmatrix}^T,\notag\\
&
C(\xi_{k})=
\begin{bmatrix}
0 & 1 & 0 & 0 & 0 & 0\\
0 & 0 & 0 & 1 & 0 & 0\\
0 & 0 & 0 & 0 & 0 & 1
\end{bmatrix},
\ \ 
D(\xi_k)=0,\label{eq:sysmodel-coef}
\end{align}
where $\xi_k=[\xi_{1k},\xi_{2k},\xi_{3k}]^T$ represents the unknown part
of the model;
$\xi_{1k}$, $\xi_{2k}$ and $\xi_{3k}$ are time-varying parameters
introduced correspondingly to $0.7s_{1k}$, $s_{2k}$ and 0.6, respectively.
With such a system model, we consider stabilizing the plant
(\ref{eq:co-state}) and (\ref{eq:co-output})
through our approach.

We first use our modeling method discussed in
Subsection~\ref{ssc:est} for determining the distribution of
$\xi_k$ in
the system model.
The simulation model for the EnKF is given by (\ref{eq:psi-x-xi})--(\ref{eq:h-C-D})
with the coefficients (\ref{eq:sysmodel-coef}) of the system model.
We temporarily regard $u$ as the input for identification at this
modeling stage, and use a maximal length sequence for the input.
In addition, we take $M=300$, $R=0.01$ and $Q=0.01{\rm
diag}(1,1,1,1,1,1,2,2,2)$ for EnKF parameters.
Then, in our computation, we obtained the response shown in Fig.~\ref{fig:estimation}
for a sample path of $s_1$ and $s_2$, where $q_0=[1,1,1,1,1,1]^T$ and
$\psi_0=[x_0^T,\xi_0^T]^T=0$ were used as the initial values.
Since the response of the states did not
diverge, the plant seems not to be unstable.
With the data of the filtered response from $k=20$ through $k=100$
(i.e., $K_0=20$ and $K_1=80$), we
obtained an empirical distribution of $\xi_k$.
The corresponding sample mean was
\begin{align}
& [0.4403, 0.1739, 0.5003]^T \label{eq:ex-mean}
\end{align}
and the covariance was
\begin{align}
&
\begin{bmatrix}
   0.0192 &  -0.0039 &   0.0081\\
   -0.0039 &   0.0750 &  -0.0006\\
    0.0081 &  -0.0006 &   0.0205
 \end{bmatrix}.\label{eq:ex-covar}
\end{align}
We can see that the above sample mean is not very far from the expected
value $[0.56, 0.2, 0.6]^T$ of $[0.7s_{1k},
s_{2k}, 0.6]^T$, and a similar comment also applies to the sample variance of
$\xi_{2k}$, which corresponds to $s_{2k}$ ($0.0750$ is not very far from $0.3^2=0.09$).
Compared to these values, the sample covariance of $[\xi_{1k},
\xi_{3k}]^T$ seems not close to the covariance of $[0.7s_{1k}, 0.6]^T$.
This might imply that the information obtained from the output of
the plant is not sufficient for distinguishing the values of
$\xi_{1k}$ and $\xi_{3k}$ in the simulation model, which may be coupled
in a single subsystem part
(i.e., the $G_2$ part).
Our purpose of obtaining the distribution of $\xi_k$ is, however, not to
obtain the exact model of the plant but 
to achieve high performance in control by taking account of
randomness behind the plant.
Fortunately, the predicted\slash filtered $x_k$ in
Fig.~\ref{fig:estimation} seems to follow
$q_k$ with a high degree of accuracy, and
the above
modeling error might not be a problem in characterizing the randomness.
Indeed, the system model with the above mean and covariance leads us to a
successful result at the next synthesis stage.

\begin{figure}[t]
  \centering
  \includegraphics[width=1\linewidth]{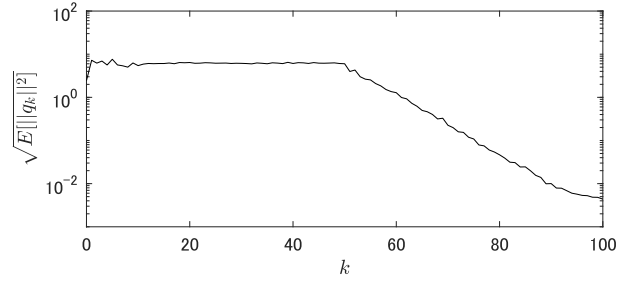}
  \caption{Response of state of plant in the closed-loop
system with our controller calculated with 200 sample paths.}
  \label{fig:decay_proposed}
\end{figure}

\begin{figure}[t]
  \centering
  \includegraphics[width=1\linewidth]{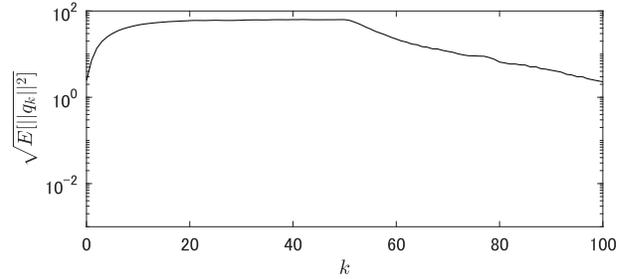}
  \caption{Response of state of open-loop plant calculated with 200 sample paths.}
  \label{fig:decay_open}
\end{figure}

With the above system model, we calculated $\bar{G}^\prime_A$ and
$\bar{G}^\prime_B$ in (\ref{eq:def-GpA})--(\ref{eq:equiv-rep-decom-syn})
(recall (\ref{eq:sample-mean})), and solved (\ref{eq:lmi-syn}) so that the
obtained state feedback gain $F=YX^{-1}$ can achieve a minimal $\lambda$
satisfying (\ref{eq:exp-def}) for the {\it ideal} closed-loop system
(\ref{eq:closed-loop}).
Then, we obtained ($\bar{n}=4$,) $\lambda=0.8432$ and 
\begin{align}
F=&[-2.4986, -4.8035, -3.8835, 2.2159,\notag\\
& 5.2592, 4.6178].
\end{align}
With this gain, we
constructed the controller described in
Fig.~\ref{fig:wcontroller},
and applied it to the present plant.
Then, 
for the external input
\begin{align}
& d_k = 
\begin{cases}
    10 & (0\leq k <50) \\
    0 & (50\leq k)
  \end{cases}
\end{align}
added on $u_k$, we obtained the response of the state $q_k$ of the plant
as shown in Fig.~\ref{fig:decay_proposed}, where this response was
calculated through approximating the expectation by the sample
mean obtained with 200 sample paths of $s_1$ and $s_2$ (as well as those
of $v^{(i)}$ and $w^{(i)}$
in the EnKF).
The EnKF parameters and the initial values were the same as those in the above modeling.
As we can see, the state after $k=50$ successfully decreases.
Although the decay rate after $k=90$ is not low compared to that before
$k=90$, this would be an effect of the artificially introduced
additive noises $v^{(i)}$ and $w^{(i)}$ in the EnKF.
If we do not use any feedback control for the present plant,
the response of $q_k$ is as shown in Fig.~\ref{fig:decay_open},
where the same environment and condition (including the external input and
samples) as the above closed-loop case was used for this computation.
It is obvious from these results that our control approach successfully
achieved the improvement of the degree of stability (in the sense of the
convergence rate of $\sqrt{E[||q_k||^2]}$).
The
convergence rate of $\sqrt{E[||q_k||^2]}$ calculated with the date at
$k=60$ and $k=80$ in Fig.~\ref{fig:decay_proposed} was $0.8473$, which
is very close to our theoretical result of $\lambda=0.8432$;
possible reasons for the small gap between these values are the
influences of modeling errors, our use of an EnKF in control, and 
the sample-base approximation in response computation.

\begin{remark}
The computation of a single response (i.e., a single sample path) of
$q_k$ in our closed-loop system from
$k=0$ to $k=100$ took from 0.38\,s to 0.51\,s with MATLAB
R2017b running on a laptop
equipped with 8.00 GB RAM and Intel(R) Core(TM) i7-5600U
CPU @ 2.60 GHz.
Hence, the average computation time for one cycle of our feedback control
was less than 10\,ms for the present example.
This implies that our controller with an EnKF could work in various actual problems
without high-spec computers.
(Although we used $M=300$ as the size of ensemble, this size is
actually conservative and can be
reduced for the present example.)
\end{remark}

\subsection{Comparison with Deterministic Systems Control Approaches
Combined with EnKF}

In the preceding subsection, we confirmed that our control approach can
improve the control performance (although only stability degree was
dealt with at this moment).
This was actually achieved by our appropriately taking account of
randomness behind the plant in our synthesis using
Theorem~\ref{th:syn}.
To further confirm this, we next compare our control approach with those
using the EnKF and conventional deterministic systems control methods.
That is, we consider the cases where the feedback gain in (\ref{eq:sf-enkf}) is designed with
other conventional methods.

If we discard the information on the covariance (\ref{eq:ex-covar}) and
take the standpoint that
the mean (\ref{eq:ex-mean}) is the true constant value of $\xi_k$ in
(\ref{eq:sysmodel-coef}), then the system model with the corresponding
coefficient matrices becomes a standard linear time-invariant (LTI) system.
Hence, we can use various conventional design methods for the LTI
system.
As an example, we first consider using the pole placement technique.
The design result of achieving the poles $(0.9, 0.8, 0.7, 0.6, 0.5, 0.4)$
is given by
\begin{align}
 F=&[-1.5556, 4.9645, 12.1698, -16.8991,\notag\\
& -71.3356, 24.6177].
\end{align}
It is obvious from the poles that this gain can stabilize the {\it ideal}
closed-loop system (\ref{eq:closed-loop}) with the above LTI system.
However, it failed to stabilize the actual closed-loop system with
the structure in Fig.~\ref{fig:wcontroller} (despite the stable behavior of the
open-loop plant in Fig.~\ref{fig:decay_open}).
A similar comment also applies to the cases with other poles in our
computation.
Hence, the pole placement technique cannot be used for the present
example.
One might consider this would be because of our using not
Theorem~\ref{th:syn} (i.e., Theorem~\ref{th:lyap})
but other methods for the LTI system;
indeed, Theorem~\ref{th:syn} can be actually used also for LTI systems
as a special case.
However, the influence of discarding the information on the
covariance is fatal even in this direction;
we obtained with Theorem~\ref{th:syn} a feedback gain
with $\lambda=0.1143$ (which is optimal for the {\it ideal}
LTI closed-loop system) that failed to stabilize the actual closed-loop system.
As is confirmed from these results, using only the mean values of the
parameters in the system model may lead us to false results for systems
with stochastic dynamics.

To circumvent this kind of issues, robust control methods
\cite{Zhou-book,rromulocconf} have been
conventionally studied, in which variations of parameters such as those in
Fig.~\ref{fig:estimation} are viewed as deterministic uncertainties.
However, the usual use of such methods in the present example is
actually impossible because the variation range of the parameters is
too large; this makes it difficult to obtain a gain that is
theoretically guaranteed to stabilize even the {\it ideal} closed-loop system.
If the variation range of the parameters used for the synthesis is narrowed, 
one could obtain some feedback gain.
However, such synthesis has an issue similar to that in the
above nominal stabilization case (the closed-loop system is unstable or
the convergence rate is not improved).
This difficulty in robust control becomes serious as the number of the
parameters to be estimated increases.

\section{Conclusions}

In this paper, we extended our state feedback control approach for
random dynamical systems toward output feedback control using an EnKF.
The EnKF was used also for modeling the distributions behind the systems.
The results of our numerical experiments demonstrated the potentiality of our
approach in improving the control performance through taking account of
the randomness behind the systems.
In particular, it might be surprising that the theoretical guarantee on
the control performance
obtained for the {\it ideal} closed-loop system (i.e., without an EnKF)
provided us with a good
estimate of that for the actual closed-loop system using the EnKF even in the
situation where the given model of the plant was partially
unknown.
Although we directly dealt with a random dynamical system as the
plant in numerical experiments, our approach is expected to
be widely effective in practical problems in which the estimates of
parameters fluctuate more than we can disregard.

Since the distribution modeling using the EnKF was successful (in the sense
of achieved control performance),
and since an estimate of the unknown parameters can also be obtained during the feedback
control (recall (\ref{eq:psi-x-xi-i})), we might be able to exploit the estimate to adjust the feedback
gain online.
That is, a sort of learning type control could be realized in which 
the controller estimates the distribution behind the
system online and determines the internal feedback gain with the estimate by itself.
Our earlier results \cite{Hosoe-TAC18,Nagira-CDC15} using random
polytopes \cite{Hug-bookchap13} would be useful for this direction of
studies; random polytopes can describe, e.g., the variation in mean and
variance of random matrices.
The present paper actually also has the role of the first step toward such advanced control.

\ifCLASSOPTIONcaptionsoff
  \newpage
\fi



\bibliographystyle{IEEEtran}
\bibliography{ms}

\vfill


\end{document}